\documentstyle[12pt,epsfig]{article}
\newcounter{figureno}
\newenvironment{capt}{
\phantom{mmmm}
\vspace*{10mm}
\parindent=0pt
\addtocounter{figureno}{1}
\begin{minipage}[t]{150mm}
\small\sl Figure~\thefigureno.\ }{\end{minipage}
\vspace*{0mm}}
\begin{document}
\thispagestyle{empty}
\pagestyle{empty}
\renewcommand{\thefootnote}{\fnsymbol{footnote}}

\renewcommand{\thanks}[1]{\footnote{#1}}
\renewcommand{\thepage}{\arabic{page}}
\catcode`@=11
\def\citer{\@ifnextchar [{\@tempswatrue\@citexr}{\@tempswafalse\@citexr[]}}

%

\def\@citexr[#1]#2{\if@filesw\immediate\write\@auxout{\string\citation{#2}}\fi
  \def\@citea{}\@cite{\@for\@citeb:=#2\do
    {\@citea\def\@citea{--\penalty\@m}\@ifundefined
       {b@\@citeb}{{\bf ?}\@warning
       {Citation `\@citeb' on page \thepage \space undefined}}%
\hbox{\csname b@\@citeb\endcsname}}}{#1}}
\catcode`@=12
\newcommand{\beq}{\begin{equation}}
\newcommand{\eeq}{\end{equation}}
\newcommand{\beqa}{\begin{eqnarray}}
\newcommand{\eeqa}{\end{eqnarray}}
\newcommand{\nn}{\nonumber}

\newcommand{\dd}{\mbox{{\rm d}}}
\newcommand{\mH}{m_{\rm H}}
\newcommand{\dLips}{\mbox{{\rm dLips}}}

\def\Im{\mbox{\rm Im\ }}
\def\Re{\mbox{\rm Re\ }}
\def\fourth{\textstyle{1\over4}}
\def\gsim{\mathrel{\rlap{\raise 1.5pt \hbox{$>$}}\lower 3.5pt
\hbox{$\sim$}}}
\def\lsim{\mathrel{\rlap{\raise 1.5pt \hbox{$<$}}\lower 3.5pt
\hbox{$\sim$}}}
\def\GeV{{\rm GeV}}
%
\def\Month{\ifcase\month\or
January\or February\or March\or April\or May\or June\or
July\or August\or September\or October\or November\or December\fi}
\def\slash#1{#1 \hskip -0.5em /}
%

\begin{flushright}
{\tt
\hfill
University of Bergen, Department of Physics \\
Scientific/Technical Report No.\ 1995-12 \\ ISSN~0803-2696\\
hep-ph/9508223 \\
July, \the\year
}
\end{flushright}
\vspace*{1cm}

\vfill
\begin{center}
{\LARGE
Determining the CP of a Higgs Particle
at a Future Linear Collider\thanks{Contributed paper,
{\em International Europhysics Conference on High Energy Physics},
July~27 -- August~2 1995, Brussels, Belgium, and
{\em 17th International Symposium on Lepton-Photon Interactions},
August 10--15, 1995, Beijing, China}}

\vspace{4mm}
{\Large
Arild Skjold and Per Osland \\\hfil\\
\vspace{3mm}
Department of Physics\thanks{Electronic mail addresses:
{\tt \{skjold,osland\}@fi.uib.no}}\\
University of Bergen \\ All\'egt.~55, N-5007 Bergen, Norway
}
\end{center}
\date{}

\vspace{10mm}
In a more general electroweak theory, there could be
Higgs particles that are odd under $CP$.
Correlations among  momenta of the initial electron and final-state
fermions are in the Bjorken process sensitive to the $CP$ parity.
Monte Carlo data on the expected efficiency demonstrate that it
should be possible to verify the scalar character of an intermediate-mass
Standard Model Higgs boson
after three years of data taking at a future linear collider.
This is most likely not possible at LEP2.
Signals of possible presence of $CP$
violation in the Higgs sector are briefly discussed.

\vfill
\newpage\normalsize
  \pagestyle{plain}
  \setcounter{footnote}{0}
  \renewcommand{\thefootnote}{\arabic{footnote}}
\section{Introduction}
\label{sec:intro}
One of the main purposes of accelerators being planned and built
today, is to elucidate the mechanism of mass generation.
In the Standard Model mass is generated via an $SU(2)$ Higgs
field doublet, associated with the existence of a Higgs particle,
whereas in more general models there are typically several
such Higgs fields, and also more physical particles.

When some candidate for the Higgs particle is discovered it
becomes imperative to establish its properties, other than the mass.
While the Standard Model Higgs boson is even under $CP$,
extended models may include pseudoscalar Higgs bosons. An example of such a
theory is the minimal supersymmetric model ($MSSM$) \cite{HHG}, where
there is a neutral $CP$-odd Higgs boson, often denoted $A^{0}$ and sometimes
referred to as a pseudoscalar.

In the context of Higgs production via the Bjorken
mechanism~\cite{Bjorken},
we have recently~\cite{skjosl3} investigated how angular distributions
may serve to disentangle
a scalar Higgs candidate from a pseudoscalar one. In trying to probe the
uniqueness of the scalar character of the Higgs boson as provided by the
Standard Model, we have to confront its predictions with those
provided by possible extensions of the Standard Model.
There is also the possibility that $CP$ violation may be present
in the Higgs sector, as first pointed out by Weinberg~\cite{Wein76}.
We briefly discuss some
possible signals of such effects.

Below we postulate an effective Lagrangian which contains $CP$ violation
in the Higgs sector.
In cases considered in the literature, $CP$ violation usually
appears as a one-loop effect. This is due to the fact that the $CP$-odd
coupling introduced below
is a higher-dimensional operator and in renormalizable models these are
induced only at loop level. Consequently we expect the
effects to be small and the confirmation of presence of $CP$ violation
to be very difficult.
Although there may be several sources of $CP$ violation, including the CKM
matrix \cite{CKM},
we will here consider a simple model where the $CP$ violation is restricted
to the Higgs sector and in particular to the coupling
between some Higgs boson and the vector bosons.
Specifically, by assuming that the coupling between the Higgs boson $H$
and the $Z$ has both scalar and pseudoscalar components,
the most general coupling for the $HZZ$-vertex
relevant for the Bjorken process may be written as
\cite{Nel,Cha}
\beq
i\ 2^{5/4} \sqrt{G_{\rm F}}
\left[ m_Z^2 \ g^{\mu \nu}
+ \xi\left(k_1^2,k_2^2\right)
\ \left(k_{1} \cdot k_{2} \ g^{\mu \nu} - k_{1}^{\mu} k_{2}^{\nu} \right)
+ \eta\left(k_1^2,k_2^2\right)
\ \epsilon^{\,\mu \nu \rho \sigma} k_{1 \rho} k_{2 \sigma} \right],
\label{EQU:int1}
\eeq
with $k_{j}$ the vector boson momentum, $j=1,2$.
The first term is the familiar $CP$-even $Z^\mu Z_\mu \ H$ tree-level
Standard Model coupling.
The second term stems from the dimension-5 $CP$-even operator
$Z^{\mu\nu} Z_{\mu\nu} H$ with
$Z_{\mu \nu}=\partial_\mu Z_\nu -\partial_\nu Z_\mu$. The last term is $CP$
odd and originates from the dimension--5 operator
$\epsilon^{\,\mu \nu \rho \sigma} Z_{\mu \nu} Z_{\rho \sigma} H$.
Simultaneous presence of $CP$-even and $CP$-odd terms leads to $CP$ violation,
whereas presence of only the last term describes a pseudoscalar
coupling to the vector bosons.
The higher-dimensional operators could be radiatively induced
and are likely to be small.

Related studies on how to discriminate $CP$ eigenstates
have been reported by \citer{Nel,HagiwaraStong}.

We demonstrate that a cut on the angle between the outgoing $Z$
and the incoming electron alters the azimuthal angular distributions
significantly \cite{skjosl3}.
Our study also addresses the problem of separating the signal from
the background.
Furthermore, we include Monte Carlo data, exploiting
the background study by Barger et.~al.~\cite{Kniehl},
and demonstrate that {\it it should be possible to verify
the scalar character of the Standard Model Higgs after
three years of running at a future linear collider}.
We also study the energy spectrum of one of the final-state
fermions in the Bjorken process, as recently suggested \cite{Arens}
in connection with Higgs decay via vector bosons to four fermions.
We compare the relative usefulness of the angular and energy
distributions.

\section{Distinguishing $CP$ eigenstates}
\label{sec:cpeig}

We compare the production of a Standard-Model Higgs ($h=H$) with
the production of a `pseudoscalar' Higgs particle ($h=A$) via the Bjorken
mechanism,
\beq
e^{-}\left(p_{1}\right) e^{+}\left(p_{2}\right)
\rightarrow Z\left(Q\right) {h}\left(q_{3}\right)
\rightarrow f\left(q_{1}\right) {\bar f\left(q_{2}\right)}
{h}\left(q_{3}\right).
\label{EQU:Bj1}
\eeq
The couplings of $H$ and $A$ to the vector bosons are given by retaining
only the first and last term in (\ref{EQU:int1}), respectively.

Let the momenta of the two final-state fermions and the initial electron
(in the overall {\it c.m.}\ frame)
define two planes,
and denote by $\phi$ the angle between those two planes; i.e.
\beq
\cos \phi = \frac{\left({\bf {p}_{1}} \times {\bf {Q}}\right)
            \cdot \left({\bf {q}_{1}} \times {\bf {q}_{2}}\right)}
                      {|{\bf {p}_{1}} \times {\bf {Q}}|
                       |{\bf {q}_{1}} \times {\bf {q}_{2}}|}.
\label{EQU:Dj4}
\eeq
We shall discuss the angular distribution of the cross section
$\sigma$,
\beq
\frac{1}{\sigma}\:
\frac{\dd\sigma}{\dd\phi}
\label{EQU:intro1}
\eeq
both in the case of $CP$-even and $CP$-odd Higgs bosons.

\begin{figure}[thb]
\begin{center}
\setlength{\unitlength}{1cm}
\begin{picture}(16,8.5)
\put(3.5,-2)
{\mbox{\epsfysize=12cm\epsffile{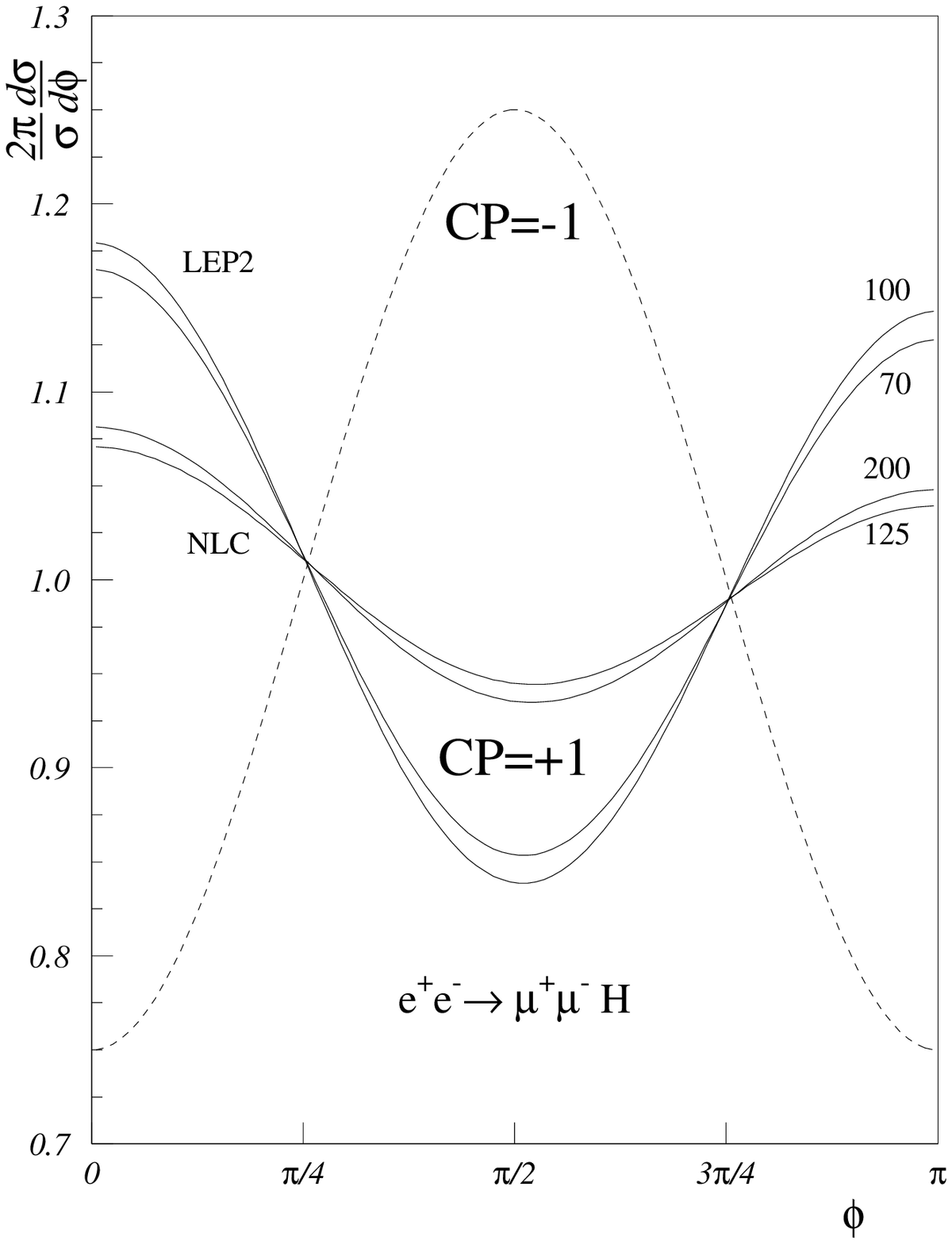}}}
\end{picture}
\begin{capt}
\label{bjlepnlc}
Angular distributions of the planes defined by
incoming $e^-$ and final-state fermi\-ons for a $CP$-even Higgs particle
(solid) compared with the corresponding distribution for a
$CP$-odd one (dashed).
Different energies and masses are considered in the $CP$-even case.
We assume $\sqrt{s}=200$ and 500~GeV at LEP2 and NLC, respectively.
\end{capt}
\end{center}
\end{figure}

The fermion-vector couplings are given by $g_V$ and $g_A$.
As a para\-meterization of these, we define
the angle $\chi$ by $g_{V} \equiv g \cos \chi$ and $g_{A} \equiv g \sin \chi$.
In the present work, the only reference to this angle
is through $\sin2\chi$
(see table~1 of ref.~\cite{osskj}).
The distributions of eq.~(\ref{EQU:intro1}) take the form
\beqa
\frac{2 \pi}{\sigma_H}\:\frac{\dd\sigma_H}{\dd\phi}
&=&
1 + \alpha(s,m) \cos \phi
  + \beta(s,m) \cos 2\phi,
\label{EQU:Dl50}
\\
\frac{2 \pi}{\sigma_A}\:\frac{\dd\sigma_A}{\dd\phi}
&=&
1 -\frac{1}{4} \cos 2\phi.
\label{EQU:Dl51}
\eeqa
The coefficients $\alpha(s,m)$ and $\beta(s,m)$ given in~\cite{skjosl3} will
at very high energies vanish as $s^{-1/2}$ and $s^{-1}$, respectively.
Therefore, the Standard-Model distribution (\ref{EQU:Dl50}) will
asymptotically at high energies become flat, whereas the
$CP$-odd distribution in eq.~(\ref{EQU:Dl51}) is independent
of energy and Higgs mass.
A representative set of angular distributions is given
in fig.~\ref{bjlepnlc} for the case
$e^+e^- \rightarrow \mu^+\mu^-h$ for both LEP2 and higher energies,
and for different Higgs masses.
There is seen to be a clear difference between the
$CP$-even and the $CP$-odd cases.

Experimentally, however, one faces the challenge of contrasting
two angular distributions with a restricted number of events
and allowing also for background.
We shall here focus on the intermediate Higgs mass range;
more specifically, we consider $m\lsim 140$~GeV where the Higgs
decays dominantly to $b\bar{b}$.
The main background will then stem from
$e^+e^- \rightarrow ZZ$ and also
$e^+e^- \rightarrow Z\gamma, \gamma \gamma$.
The cleanest channel for isolating the Higgs signal from the background
is provided by the $\mu^+\mu^-$ and
$e^+ e^-$ decay modes of the $Z$ boson.

Let us next limit consideration to the energy range
$\sqrt{s}=300-500$~GeV, as appropriate for a linear collider
\cite{Wiik}, henceforth denoted NLC.
We impose reasonable cuts and constraints as described in
\cite{Kniehl}; e.g. $|m_{\mu^+\mu^-}-m_Z| \leq 6$~GeV and
$|\cos\theta_Z| \leq 0.6$.
The signal for
$e^+e^- \rightarrow Z H \rightarrow \mu^+\mu^- b \bar{b}$ will then
be larger than the background
$e^+e^- \rightarrow Z Z \rightarrow \mu^+\mu^- b \bar{b}$
by an order of magnitude.
In the following we shall thus neglect the background
in the discussion of (\ref{EQU:Dl50}) versus (\ref{EQU:Dl51}).
With  $\sigma(e^+e^- \rightarrow Z H) \sim 200$~fb and an integrated
luminosity of 20 ${\rm fb}^{-1}$ a year \cite{Kniehl},
about 4000 Higgs particles will be produced per year, in this
intermediate mass range.
However, following~\cite{Kniehl} we have only $\sim 30$ signal events
$e^+e^- \rightarrow Z H \rightarrow \mu^+\mu^- b \bar{b}$ left
per year for e.g.\ a NLC operating at $\sqrt{s}=300$~GeV
and a Higgs particle of mass $m=125$~GeV.
In the case $e^+e^- \rightarrow Z H \rightarrow e^+ e^- b \bar{b}$
we also have a t-channel background contribution
from the $ZZ$ fusion process
$e^+e^- \rightarrow e^+e^- (Z Z) \rightarrow e^+ e^- H$.
This contribution may be neglected at LEP energies,
but it is comparable to the s-channel contribution at higher
energies. However, this contribution can be suppressed
by imposing a cut on the invariant mass of the final-state electrons,
e.g.\ $|m_{e^+e^-}-m_Z| \leq 6$~GeV.
Hence, we can effectively treat the electrons on the same footing as
the muons, thereby obtaining a doubling of the event rate.

Imposing the cut $|\cos\theta_Z| \leq b$, the predictions for
the azimuthal correlations of eqs.~(\ref{EQU:Dl50})--(\ref{EQU:Dl51})
get modified. For the $CP$-even case we find~\cite{skjosl3}
\beqa
\alpha^{b}(s,m)
&=          &
\sin2\chi\, \sin2\chi_1
\left(\frac{3\pi}{4}\right)^2\,
\frac{\sqrt{s}\, m_Z \left(s+m_Z^2-m^2\right)}
{\xi(b)\lambda\left(s,m_Z^2,m^2\right)+12 s\, m_Z^2}
\,\zeta(b),
\nn \\
\beta^{b}(s,m)
&=          &
\frac{2\xi(b)s\, m_Z^{2}}
{\xi(b)\lambda\left(s,m_Z^{2},m^{2}\right)+12 s\, m_Z^{2}},
\label{EQU:El53}
\eeqa
where $\lambda(x,y,z)=x^2+y^2+z^2-2(xy+xz+yz)$ is the K{\"{a}}llen function and
\beqa
\xi(b)
&=          &
\frac{1}{2}\left(3-b^2\right), \qquad \xi(1)=1, \nn \\
\zeta(b)
&=          &
\frac{2}{\pi}
\left(\frac{\frac{\pi}{2}-\arccos b}{b}+\sqrt{1-b^2}\right),
\qquad \zeta(1)=1,
\eeqa
whereas for the $CP$-odd case
\beqa
-\frac{1}{4}
&\rightarrow&
-\frac{\xi(b)}{3+b^2}.
\label{EQU:El54}
\eeqa

\begin{figure}[t]
\begin{center}
\setlength{\unitlength}{1cm}
\begin{picture}(16,8.5)
\put(0.2,-2){\mbox{\epsfysize=11cm\epsffile{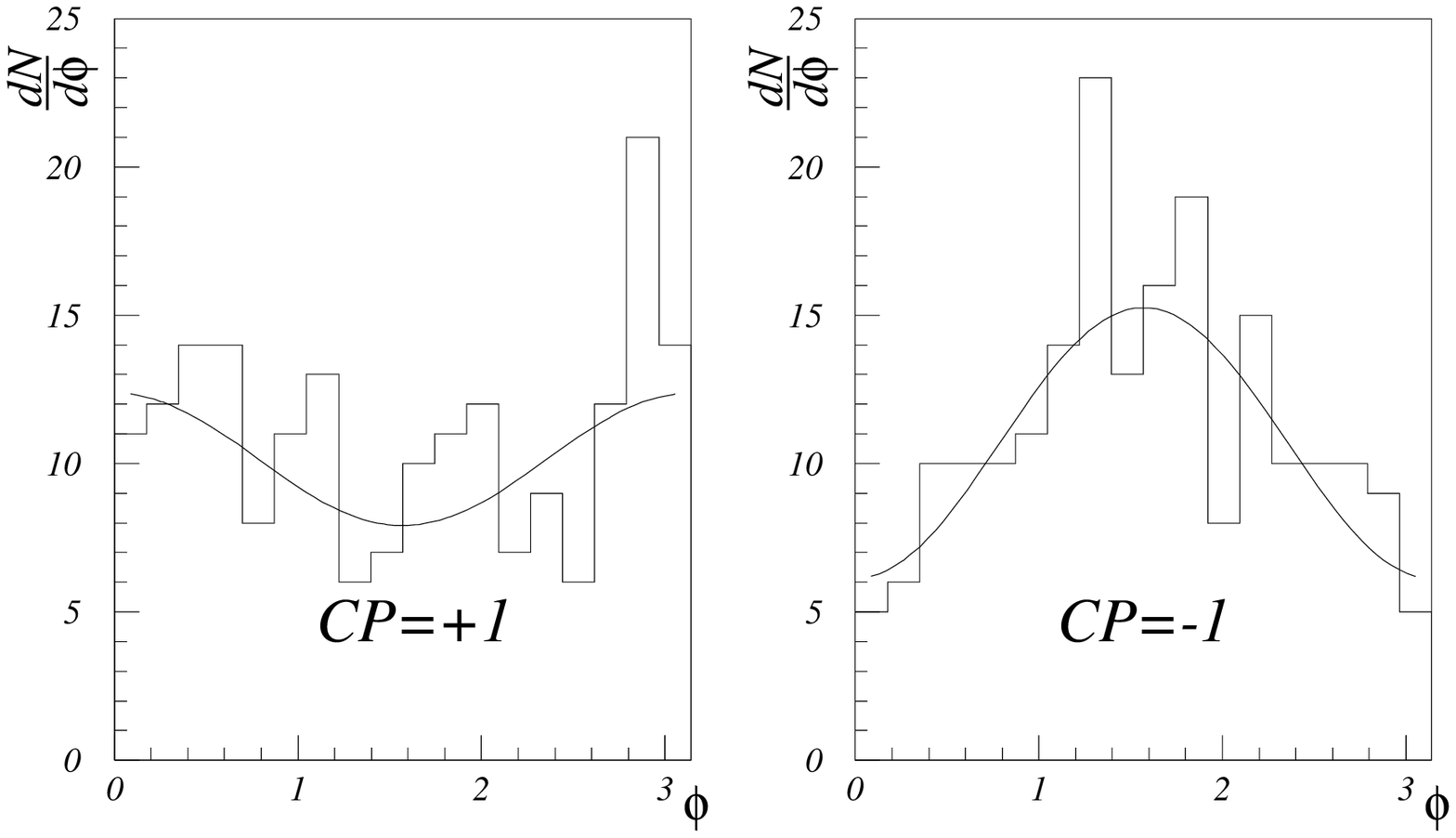}}}
\end{picture}
\begin{capt}
\label{bjmcnlc}
Monte Carlo data
displaying the angular distribution of events
$e^+e^- \rightarrow Z H \rightarrow l^+l^- b \bar{b}$, $l=\mu,e$
for a Standard-Model Higgs versus a CP-odd one.
We have taken $\sqrt{s}=300$~GeV, $m=125$~GeV, and an angular cut
$|\cos\theta|\le b=0.6$.
\end{capt}
\end{center}
\end{figure}

In order to demonstrate the potential of the NLC for determining
the $CP$ of the Higgs particle,
we show in fig.~\ref{bjmcnlc} the result of a Monte Carlo simulation.
For this purpose we have used PYTHIA \cite{Sjostrand},
suitably modified to allow for the $CP$-odd case.
The statistics correspond to 3~years of running\footnote{The event
rate is based on the Standard Model, and could be different for
a non-standard Higgs sector.}
using both the $\mu^+\mu^-$ and $e^+ e^-$ decay modes of the $Z$ boson.
This yields about 200 events in these channels.
Although the cut $b=0.6$ makes $\alpha$ increase as shown in
(\ref{EQU:El53}), the $\cos\phi$ term is still too small to show up
in the Monte Carlo simulation.
For $\sqrt{s}=300~\GeV$ and $m_H=125~\GeV$,
the `bare' prediction (\ref{EQU:Dl50}) for $\beta$ is 0.12.
The cut $b=0.6$ increases it slightly to 0.14.
Similarly, the `$-1/4$' of (\ref{EQU:Dl51})
changes significantly to $-0.39$. Consequently, the cut makes it easier
to discriminate between the $CP$-even distribution and the $CP$-odd one.
{}From fig.~\ref{bjmcnlc} we see that the individual angular
Monte Carlo distributions are consistent with the predictions,
showing that a three-year data sample is large enough to
reproduce the azimuthal distributions.
In the Standard-Model case the fit gives $0.92\pm 0.07$ and
$0.2\pm 0.1$ for the predictions 1.00 and 0.14, respectively,
with $\chi^2=1.0$.
In the $CP$-odd case the fit gives $0.94\pm 0.07$ and
$-0.4\pm 0.1$ for the predictions 1.00 and $-0.39$, respectively,
with $\chi^2=0.7$.
More importantly, since the $\cos 2\phi$ terms are more than 4 standard
deviations away, a data sample of this size is sufficient
to verify the scalar nature of the
Standard-Model Higgs.
Using likelihood ratios, as described in \cite{Roe}, for choosing between the
two hypotheses of $CP$ even and $CP$ odd, we find that less than 3 years of
running suffices using similar criteria.

\begin{figure}[thb]
\begin{center}
\setlength{\unitlength}{1cm}
\begin{picture}(16,7.5)
\put(0.,-2){\mbox{\epsfysize=11cm\epsffile{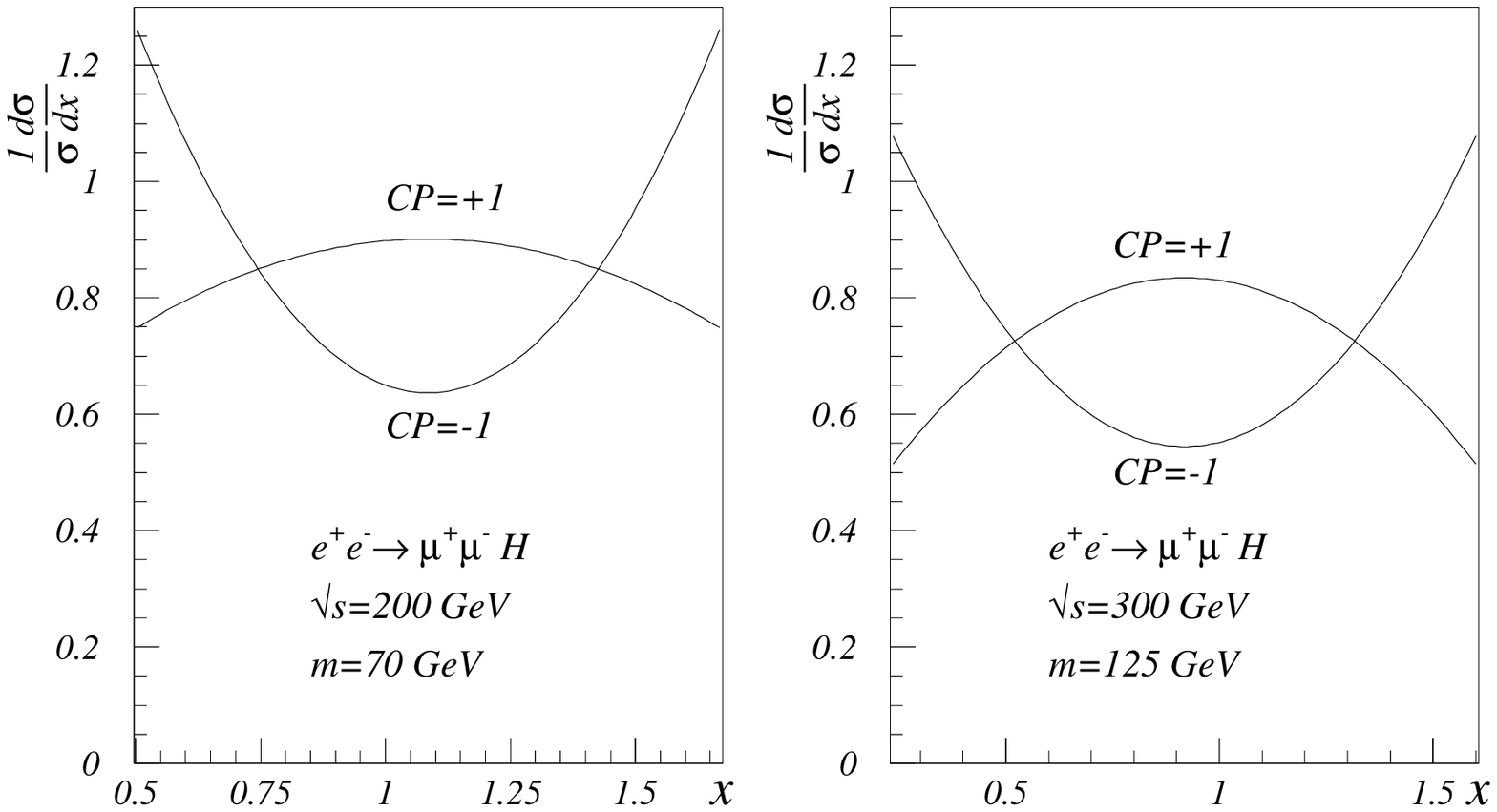}}}
\end{picture}
\begin{capt}
\label{bjspec}
Characteristic distributions
for the scaled energy of the $l^-$,
$l=\mu,e$ in the Bjorken process $e^+e^- \rightarrow l^+l^- h$.
Different energies and masses are considered.
\end{capt}
\end{center}
\end{figure}

An alternative test has recently been suggested
by Arens et.~al.~\cite{Arens} in the context of Higgs
decaying via vector bosons to four fermions, where one studies
the energy spectrum of one of the final-state fermions.
Applying this idea to the Bjorken process one would
study the energy distribution of an outgoing fermion, e.g.\ $\mu^-$ or $e^-$.
Introducing the scaled lepton energy, $x=4E_{l^-}/\sqrt{s}$, $l=\mu,e$,
we shall consider the energy distribution of the cross section
with respect to this final-fermion energy,
\beq
\frac{1}{\sigma}\:
\frac{\dd\sigma}{\dd x}
\label{EQU:intro2}
\eeq
both in the case of $CP$-even and $CP$-odd Higgs bosons. We are using
the narrow-width approximation and the range of $x$ is given by
$x_-\le x\le x_+$, with $sx_{\pm}=s+m_Z^2-m^2\pm\sqrt{\lambda}$.
Here the distributions are given as second-degree polynominals
in $x$, and, as shown in~\cite{skjosl3}, the coefficients have a non-trivial
dependence on the {\it c.m.}  energy and the Higgs mass, also for the
$CP$-odd case.
A representative set of energy distributions is given in
fig.~\ref{bjspec} for the case $e^+e^- \rightarrow \mu^+\mu^- h$
for both LEP2 and NLC energies.
There is a clear difference between the $CP$-even
and the $CP$-odd cases.
\begin{figure}[t]
\begin{center}
\setlength{\unitlength}{1cm}
\begin{picture}(16,7.5)
\put(0.8,-2){\mbox{\epsfysize=10cm\epsffile{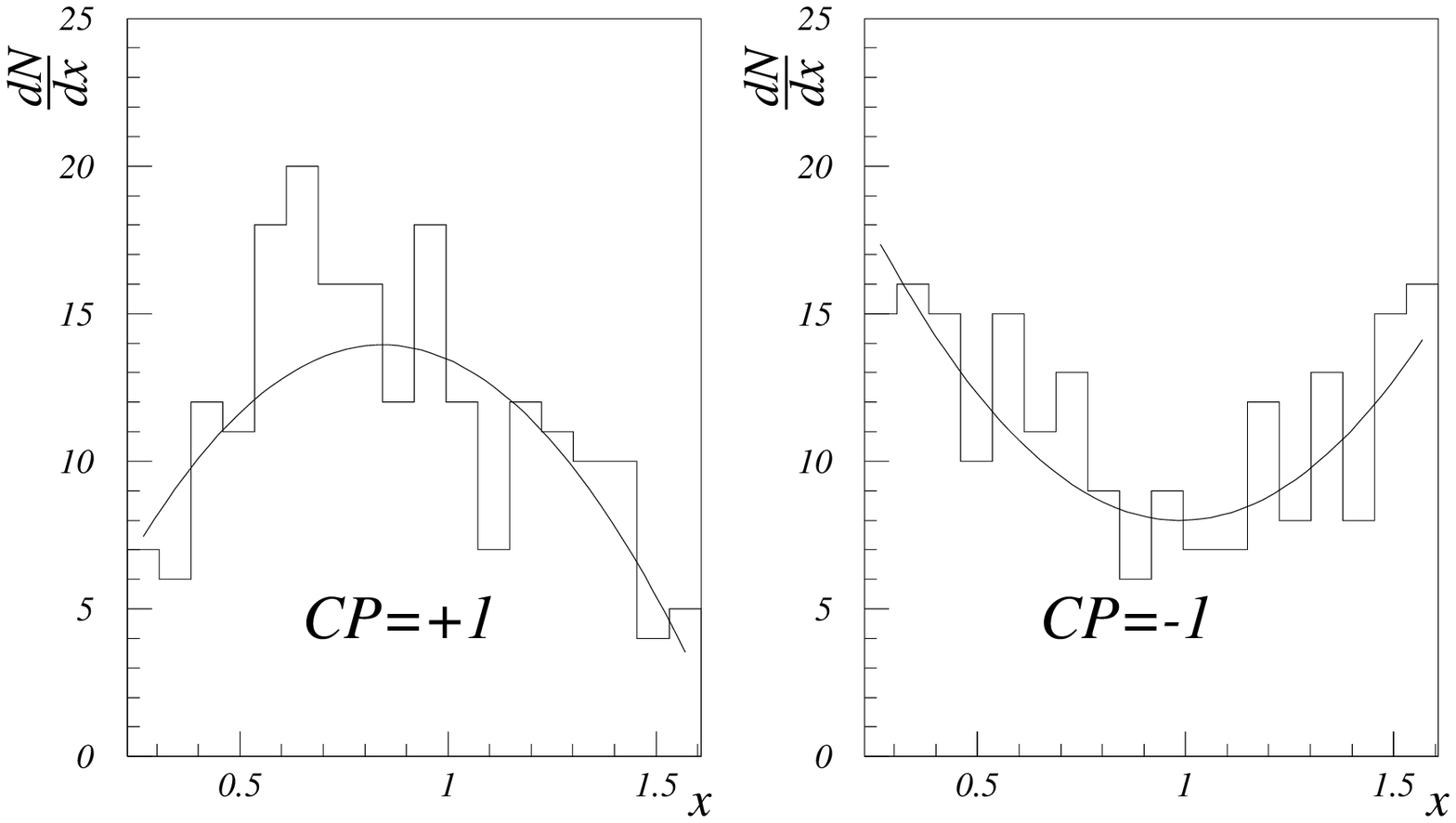}}}
\end{picture}
\begin{capt}
\label{bjmcx}
Monte Carlo data
displaying the lepton energy distribution for events
$e^+e^- \rightarrow Z H \rightarrow l^+l^- b \bar{b}$, $l=\mu,e$
for a Standard-Model Higgs versus a CP-odd one.
We have taken $\sqrt{s}=300$~GeV and $m=125$~GeV.
\end{capt}
\end{center}
\end{figure}

In fig.~\ref{bjmcx} we show the result of a Monte-Carlo simulation
for the energy distribution eq.~(\ref{EQU:intro2}) analogous
to the one in fig.~\ref{bjmcnlc}. Again, a cut on the polar angle is
imposed.
As in the case of angular distributions,
the cut makes it easier to discriminate between the
$CP$-even distribution and the $CP$-odd one.
Also in this case the data sample
reproduces the predicted energy distributions.
An analysis of the likelihood ratios demonstrates that less than
3 years of running is sufficient if we require the correct answer
with a discrimination by four standard deviations,
but more events seem to be required than in the case of angular
distributions.

\section{$CP$ violation}
\label{sec:vio}
As previously mentioned, if we allow for both the Standard-Model
and the $CP$-odd term in the Higgs-vector coupling (\ref{EQU:int1}),
then there will be $CP$ violation.
This situation will be discussed here.
It is similar to the case of Higgs decay discussed elsewhere
\cite{skjoldosland}.
We discard the higher-dimensional $CP$-even term since it is
likely to be small.

In terms of the invariant mass $s_1$ of the fermion pair,
and neglecting terms of ${\cal O}((\Im\eta)^2)$,
the distribution (\ref{EQU:intro1}) can be written compactly as
\beqa
 \frac{\dd^2\sigma }{\dd\phi\: \dd s_1}
&  =   & \frac{N_1}{144 \sqrt{2} (4\pi)^4}\, \frac{G_{\rm F}}{s^2}
\sqrt{\lambda\left(s,s_1,m^2\right)}\, D(s,s_1) \nn \\
&\times&
\biggl[ \lambda\left(s,s_1,m^2\right)+4 s s_1 \left(1+2 \rho^{2} \right)
+2s s_1\,\rho^2\,\cos 2(\phi + \delta) \nn \\
&  +   &   \sin 2 \chi \sin 2 \chi_{1} \left(\frac{3 \pi}{4}\right)^2
\sqrt{s s_1}\, (s+s_1-m^2)\,\rho\,\cos (\phi + \delta) \biggr],
\label{EQU:no3}
\eeqa
with a modulation function
\beq
\rho=\sqrt{1+\left(\Re\eta\right)^2\lambda\left(s,s_1,m^2\right)/(4m_Z^4)},
\label{EQU:rho}
\eeq
and an angle
\beq
\delta=
\arctan\frac{\Re\eta(s,s_1)\sqrt{\lambda(s,s_1,m^2)}}{2m_Z^2}, \qquad
-\pi/2 < \delta < \pi/2,
\label{EQU:delta}
\eeq
describing the relative shift in the angular distribution
of the two planes, due to $CP$ violation.
This rotation vanishes at the threshold for producing
a real vector boson (where $\lambda=0$) and,
even for a fixed value of $\Re\eta$, grows with energy
(because of the $\sqrt{\lambda}$-factor).

This relation (\ref{EQU:delta}) can be inverted to give for the
$CP$-odd term in the coupling:
\beq
\Re\eta=\frac{2m_Z^2}{\sqrt{\lambda(s,s_1,m^2)}}\, \tan\delta.
\eeq
This result (\ref{EQU:no3}) is completely analogous to the one
encountered for the decay of Higgs particles, eq.~(12) of
\cite{skjoldosland}, if we interchange $\phi$ and $\pi-\phi$.

Above threshold for producing a real vector meson accompanying the
Higgs particle, we may integrate over $s_1$ in the narrow-width
approximation.
Imposing the cut $|\cos\theta_Z| \leq b$, the distribution of
eq.~(\ref{EQU:intro1}) takes the compact form
\beq
\frac{2 \pi}{\sigma^b}\:\frac{\dd\sigma^b}{\dd\phi}
= 1 + \alpha^{b\, \prime}(s,m) \, \rho \, \cos (\phi +\delta)
+ \beta^{b\, \prime}(s,m) \, \rho^{2} \, \cos 2\left(\phi +\delta\right).
\label{EQU:Dl5}
\eeq
with $\rho$ and $\delta$ given by eqs.~(\ref{EQU:rho})
and (\ref{EQU:delta}), substituting $s_1=m_Z^2$.
The details of implementing a cut in
polar angle are given in~\cite{skjosl3}.
Any $CP$ violation would thus show up as a ``tilt'' in the
azimuthal distribution, by the amount $\delta$.

\begin{figure}[t]
\begin{center}
\setlength{\unitlength}{1cm}
\begin{picture}(16,8.5)
\put(3.5,-2){\mbox{\epsfysize=12cm\epsffile{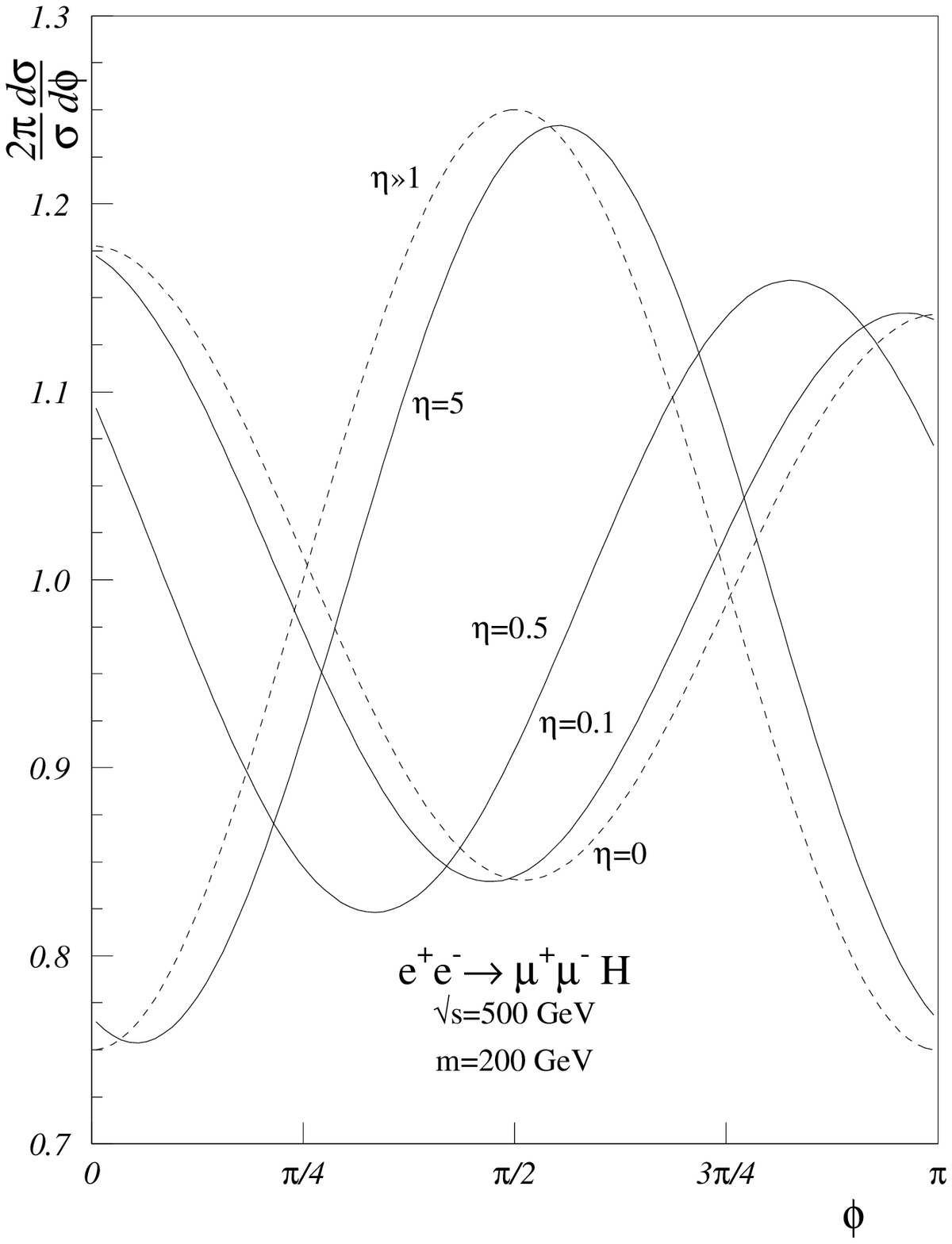}}}
\end{picture}
\begin{capt}
\label{plcp}
for different amounts of $CP$
violation, including the $CP$-even ($\eta=0$) and $CP$-odd ($|\eta|\gg1$)
eigenstates. We have used
$\Re\eta=0.1, 0.5, 5$ for $\sqrt{s}=500$~GeV and $m=200$~GeV.
\end{capt}
\end{center}
\end{figure}
A representative set of angular distributions is given in
fig.~\ref{plcp} for a broad range of $\Re\eta$ values.
We have considered a Higgs boson of $m=200$~GeV
accompanied by a $\mu^{+} \mu^{-}$-pair in the final state,
produced at $\sqrt{s}=500$~GeV.
We observe that for $\Re\eta \lsim 0.1$ and $\Re\eta \gsim 5$,
the deviations from the $CP$-even and $CP$-odd distributions,
respectively, are small.
Experimentally it will be very difficult to disentangle two
distributions which differ by such a small phase shift.
Thus, observation of a small amount
of $CP$ violation would require a very large amount of data.
This should be compared with the situation
in fig.~\ref{bjlepnlc} and fig.~\ref{bjmcnlc}.

We note that the special cases $\eta=0$ and $|\eta| \gg 1$
correspond to the $CP$ even and $CP$ odd eigenstates,
respectively. Hence, the distribution (\ref{EQU:Dl5})
should be interpreted as being intermediate between those for
the two eigenstates.

\section{Summary}
\label{sec:conc}

We have addressed the problem of estimating the amount of data needed
in order to distinguish a scalar Higgs from a pseudoscalar one
at a future linear collider.
This is most likely not possible at LEP2 due to much smaller event rates and
background which is not easily suppressed.
However, we have demonstrated that one will be able to establish
the scalar nature of the Higgs boson at the Next Linear Collider
from an analysis of angular or energy correlations.
This particular study has been carried out for the case $\sqrt{s}=300$~GeV,
$m=125$~GeV. Similar results are expected in other cases as long as
the background is small.
In cases where the background can not be significantly suppressed a more
dedicated study would be required.

In order to establish or rule out specific models, one will also need
to compare different branching ratios, in particular to fermionic
final states.  The methods proposed above instead deal with quite
general properties of the models.

This research has been supported by the Research Council of Norway.

\end{document}